\documentclass[prb,twocolumn,showpacs,amsmath,amssymb,floats,floatfix,prl,aps,superscriptaddress,showkeys]{revtex4}

\usepackage{graphicx}
\usepackage{epsfig}
\usepackage{dcolumn}
\usepackage{bm}
\usepackage{times}

\begin{document}


\title{Magnetic quantum coherence effect in Ni$_4$ molecular transistors}

\author{Gabriel Gonz\'alez}

\email{gabrielglez@iteso.mx}
\affiliation{Departamento de Matem\'aticas y F\'isica, Instituto Tecnol\'ogico y de Estudios Superiores de Occidente, \\ Perif\'erico Sur Manuel G\'omez Mor{\'i}n 8585 C.P. 45604, Tlaquepaque, Jal., MEXICO}

\author{Michael N. Leuenberger}
\email{Michael.Leuenberger@ucf.edu}
\affiliation{NanoScience Technology Center, University of Central
Florida, Orlando, FL 32826, USA,}

\affiliation{Department of Physics, University of Central
Florida, P.O. Box 162385, Orlando, FL 32816-2385, USA}

\date{\today}

\begin{abstract}
We present a theoretical study of electron transport in Ni$_4$ molecular transistors in the presence of Zeeman spin splitting and magnetic quantum coherence (MQC). The Zeeman interaction is extended along the leads which produces gaps in the energy spectrum which allow electron transport with spin polarized along a certain direction. We show that the coherent states in resonance with the spin up or down states in the leads induces an effective coupling between localized spin states and continuum spin states in the single molecule magnet and leads, respectively. 
We investigate the conductance at zero temperature as a function of the applied bias and magnetic field by means of the Landauer formula, and show that the MQC is responsible for the appearence of resonances. Accordingly, we name them MQC resonances.
\end{abstract}



\maketitle

\section{Introduction}
\label{sec:introduction}

Single-molecule magnets (SMMs), such as Mn$_{12}$ (see
Refs. \onlinecite{experiments,delBarco}) and Fe$_8$ (see
Refs. \onlinecite{Sangregorio,Wernsdorfer}), have become the focus of
intense research since experiments on bulk samples demonstrated the
magnetic quantum tunneling of a single magnetic moment on a macroscopic scale. 
These molecules are characterized by a large total spin, a large
magnetic anisotropy barrier, and anisotropy terms which allow the spin
to tunnel through the barrier. It is well known that magnetic quantum coherence (MQC) is realized when the SMM tunnels several times between degenerate spin states with opposite spin projections on the magnetic easy axis before the coherence is destroyed by the environment. Evidence of MQC has been reported for various superconducting systems and for antiferromagnetic nanoclusters. \cite{nakamura, vion, yu, awschalom,garg1,garg2}\\
Electronic transport through SMMs
offers several unique features with potentially large impact on
applications such as high-density magnetic storage as well as quantum
computing.\cite{app,lapo} Recent experiments have pointed out the
importance of the interference between spin tunneling paths in
molecules. For instance, measurements of the magnetization in bulk
Fe$_8$ have observed oscillations in the tunnel splitting $\Delta_{s,-s}$ between states $S_z = s$ and $-s$ as a function of a
transverse magnetic field at temperatures between $0.05$ K and $0.7$ K
(see Ref. \onlinecite{Zener,Gunther}). This effect can be explained by the
interference between Berry phases associated to spin tunneling path
of opposite windings.\cite{LossDelftGarg,Leuenberger_Berry,future,gabriel,gabriel1} \\
In this article we investigate coherent magnetic quantum tunneling in the SMM Ni$_4$ in which the tunneling rate is faster than the rate of decoherence and at a temperature at which tunneling occurs only between the lowest spin states. We have chosen to study the SMM Ni$_4$ because of its high symmetry (S$_4$) and large tunnel splittings ($\sim$ 0.01K) at zero magnetic field, which have been confirmed by high frequency EPR and magnetic relaxation experiments (see Ref. \onlinecite{yang,delboat}). The total spin ground state of Ni$_4$ is $S=4$.\\ 
The article is organized as follows. First we will start with the model Hamiltonian of the Ni$_4$ molecular transistor taking into account the magnetic quantum coherence of the two lowest ground spin states of the SMM. Then we will solve this model Hamiltonian and use the solution to calculate the conductance through the molecular transistor as a function of the applied bias by means of the Landauer formalism at zero temperature for the SMM Ni$_4$. The conclusions are summarized in the last section.

\section{Model Hamiltonian}
\label{sec:model}

\begin{figure}[!h]
\begin{tabular}{m{-5cm}m{18cm}}
\multicolumn{2}{c}{} \\ [-0.5cm]
\resizebox{80mm}{!}{\includegraphics{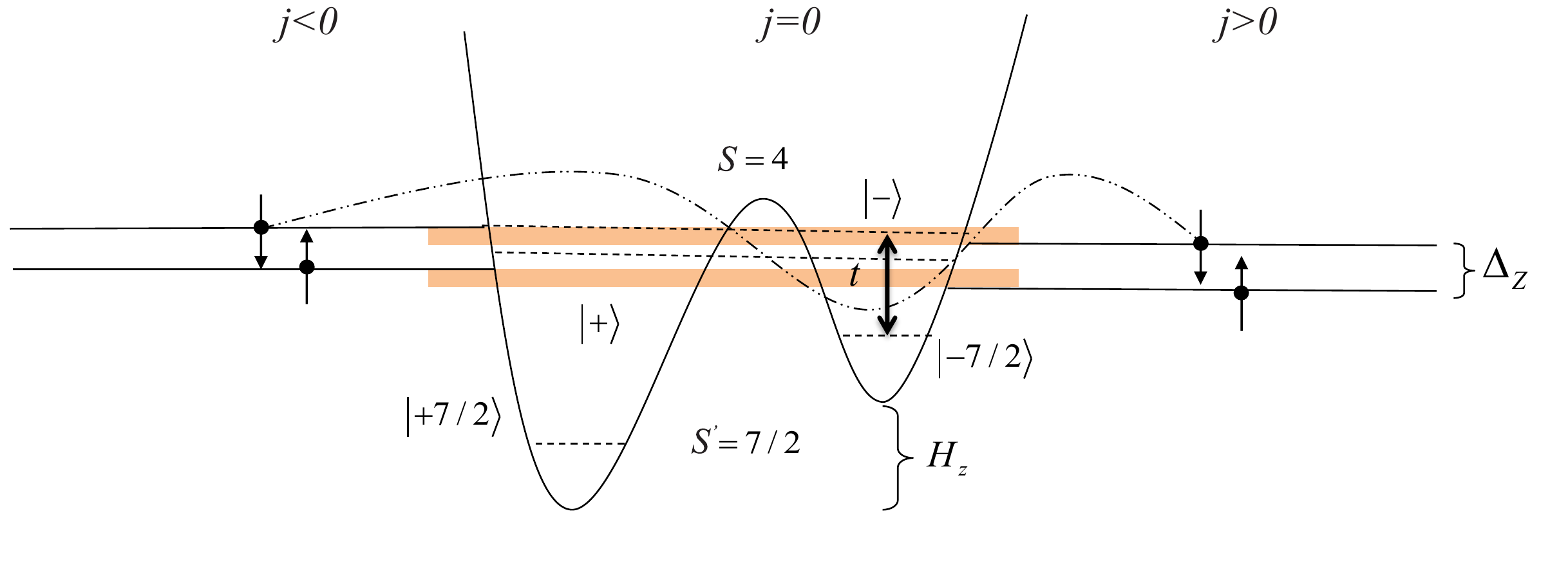}} & \vspace{1cm} \mbox{\bf (a)} \\
\resizebox{80mm}{!}{\includegraphics{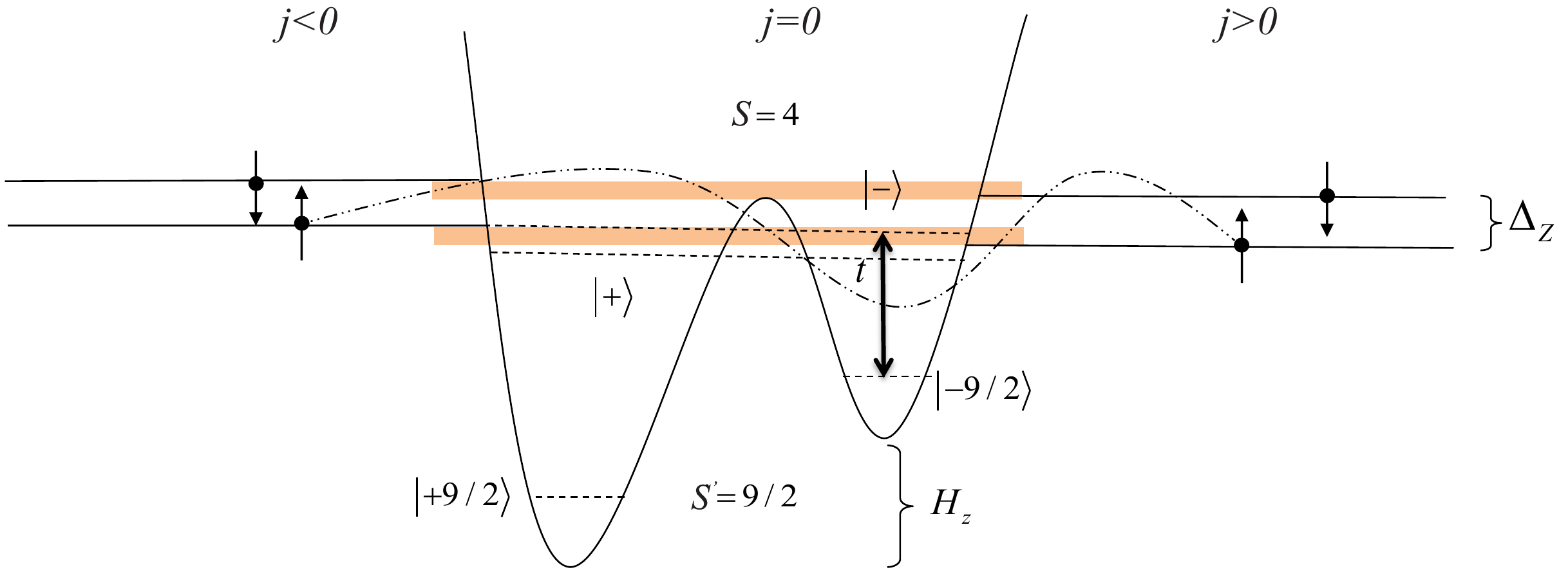}} & \mbox{\bf (b)} \\
\end{tabular}
\caption{Schematic illustration of tunneling through the SMM transistor for $s=7/2$ and $s=9/2$. The dotted line corresponds to the direct electron path across the SMM transistor.}
\label{scheme}
\end{figure}

The total Anderson-type Hamiltonian of a system formed by a single-molecule magnet
(SMM) attached to two metallic leads can be
separated into three terms
\begin{equation}
\label{eq:Htotal}
{\cal H}_{\rm tot} = {\cal H}_{\rm lead} + {\cal H}_{\rm SMM} + {\cal
H}_{\rm SMM-lead}.
\end{equation}
We will consider the leads as a one dimensional linear chain of sites. Thus, 
the first term on the right-hand side of
Eq. (\ref{eq:Htotal}) is given by,
\begin{equation}
\label{eq:Hlead}
{\cal H}_{\rm lead} =  \sum_{i}\sum_{\sigma=\uparrow,\downarrow} \epsilon_{\sigma}c^{\dagger}_{i\sigma}c_{i\sigma}-
v\sum_{\langle ij\rangle}\sum_{\sigma=\uparrow,\downarrow} \left(c^{\dagger}_{i\sigma}c_{j\sigma}+c^{\dagger}_{j\sigma}c_{i\sigma}\right),
\end{equation}
where the operator
$c_{i\sigma}^\dagger$ ($c_{i\sigma}$) creates (annihilates)
electronic states in the leads with 
spin orientation $\sigma=\uparrow,\downarrow$, and energy $\epsilon_{\sigma}$. The symbol $\langle ij\rangle$ implies the sum over nearest neighbors. The potential of the wire is taken to be
zero and the hopping in the wire is $v$. The on site energies in the leads and in the SMM are given by $\epsilon_{\sigma}=\frac{\Delta_z}{2}[\sigma_z]_{\sigma\sigma}$ and $\varepsilon_{0\sigma}$, respectively. $\Delta_z=g\mu_BH_z$ is the Zeeman energy splitting in the leads where $\sigma_z$ is the Pauli matrix.\cite{grechnev}\\
The second term on the right-hand side of Eq. (\ref{eq:Htotal}) denotes
the SMM part, which can be broken into spin, charging, and
gate contributions,
\begin{equation}
\label{eq:HSMM}
{\cal H}_{\rm SMM} = {\cal H}_{\rm spin}^{(q)} + E_c - q\, e V_g,
\end{equation}
where $E_c$ denotes the charging energy, $q$ is the number of excess
electrons (the charge state of the molecule), and $V_g$ is the
electric potential due to an external gate voltage. In the
presence of an external magnetic field, the spin Hamiltonian of the
SMM Ni$_4$ reads
\begin{eqnarray}
{\cal H}_{\rm spin}^{(q)} & = &-D_qS_{q,z}^2+C_q(S_{q,+}^4+S_{q,-}^4)-\mu_B g\vec{S}\cdot\vec{H} \nonumber \\ &=&-D_qS_{z}^2+C_q(S_{q,+}^4+S_{q,-}^4)-\frac{1}{2}
\left( h_\bot^\ast S_{q,+}+ h_\bot S_{q,-} \right) \nonumber \\ & &+ h_\parallel S_{q,z},
\label{H_spin}
\end{eqnarray}
where the easy axis is taken along the $z$ direction and $S_{q,\pm} =
S_{q,x} \pm iS_{q,y}$. The magnetic field components were rescaled to
$h_\bot = g\mu_B (H_x + iH_y)$ and $h_\parallel = g\mu_B H_z$ for the
transversal and longitudinal parts, respectively, where $g=2.3$ denotes
the effective gyromagnetic ratio for the giant spin of the SMM and $D_{q=0}=0.75 K$ and $C_{q=0}=2.9\times10^{-4} K$. The total spin as well as the anisotropy constants $D_q$ and $C_q$ depend on
the charging state of the molecule, i.e. if the SMM is singly and doubly charged.\cite{dataexps}
The longitudinal magnetic field $H_z$ tilts the double potential well favoring those spin projections aligned with the field. At zero magnetic field the spin projections $|4\rangle$ and $|-4\rangle$ have nearly the same energy and magnetic quantum tunneling is possible. Importantly, for an individual Ni$_4$ nanomagnet quantum magnetic tunneling is possible only between states that differ by 4 spin units, i.e. if the selection rule $s_{q}-s'_{q}=4k$ is satisfied, where $k$ is an integer.\cite{gatt} 
The transverse magnetic field in the $xy$ plane lifts the degeneracy of the eigenstates of $S_{q,z}$ by an energy $\Delta_{s_{q},s'_{q}}$, the so-called tunnel splitting, and leads to states that are coherent superpositions of the eigenstates of $S_{q,z}$. Denoting $\lambda_{s,-s}$ as the coupling matrix element between the states $|s\rangle_{q=0}$ and $|-s\rangle_{q=0}$, the magnetic quantum coherence in the single molecule magnet is given by the following effective Hamiltonian
\begin{eqnarray}
\label{eqmqc}
{\cal H}_{MQC} &=& \lambda_{s,-s}\left(\left|-s\rangle_{00}\langle s\right|+\left|s\rangle_{00}\langle-s\right|\right).
\end{eqnarray}
where $\lambda_{s,-s}$ represents the source of spin flipping.
The most general coherent superpositions for the two lowest levels $|s\rangle_{q=0}$ and $|-s\rangle_{q=0}$ is given by
\begin{equation}
\left|\pm\right>=\frac{\left(\mp\Delta_{s,-s}+\sqrt{\Delta^2_{s,-s}+\lambda_{s,-s}^2}\right)\left|-s\right>_0\pm\lambda_{s,-s}\left|s\right>_0}{N_{\pm}},
\label{eqsym}
\end{equation}
where 
\begin{equation}
N_{\pm}=\sqrt{\lambda^2_{s,-s}+\left(\mp \Delta_{s,-s}+\sqrt{\Delta^2_{s,-s}+\lambda^2_{s,-s}}\right)^2}.
\label{eqnorm}
\end{equation} 
Note that when $\lambda_{s,-s}\rightarrow 0$ in Eq.(\ref{eqsym}) then $|\pm\rangle\rightarrow |-s\rangle_0$, which means that there is only one conducting channel open if there is no magnetic quantum coherence.
To take advantage of the coherent superposition in Eq.(\ref{eqsym}) we need to singly charge the SMM Ni$_4$ in order to switch the total spin ground state from $S=4$ to $S=9/2$ or $S=7/2$, the final ground state will be the result of the exchange interaction between the total spins in the SMM. When the nanomagnet is singly charged and with the application of the longitudinal magnetic field $H_z$ the SMM will only allow electrons with spin down (up) polarization in single electron tunneling transport due to spin blockade. If the SMM has a total spin ground state of $S=9/2$ or $S=7/2$, then there will be transitions from $|-9/2\rangle$ to $|\pm\rangle$ and from $|-7/2\rangle$ to $|\pm\rangle$, respectively. For $S=9/2$ spin up electrons are transmitted through the SMM, whereas for $S=7/2$ spin down electrons are transmitted through the SMM. The energy levels as a function of the orientation of the magnetic moment is pictorially shown in Figure (\ref{scheme}), where the gap between $|+>$ and $|->$ has been exaggerated to help visualization of the electron transmission through the SMM transistor. \\
We will be using two pairs of spin ground states $|\pm s\rangle_{q=0}$ and $|\pm s\rangle_{q=1}$ for the uncharged and charged SMM, respectively. For the sake of clarity, we will denote the charged ground states by $|\pm s'\rangle$ and the uncharged ground states by $|\pm s\rangle$. With this notation and by restricting the Hilbert space to the lowest spin doublet of the SMM we can then write down the Hamiltonian which represents the interaction of the SMM with the leads 
\begin{eqnarray}
\label{eq:SMM-lead}
{\cal H}_{\rm SMM-lead}&=& -t \left( 
\left|\pm\right>\left<-s'\right| c_{1\downarrow} + c^{\dagger}_{1\downarrow} \left|-s'\right>\left<\pm\right| \right. +\nonumber \\
& & \left.\left|\pm\right>\left<-s'\right| c_{1\uparrow} + c^{\dagger}_{1\uparrow} \left|-s'\right>\left<\pm\right|  \right.+\nonumber \\
& & \left. 
\left|\pm\right>\left<-s'\right| c_{-1\downarrow} + c^{\dagger}_{-1\downarrow} \left|-s'\right>\left<\pm\right| \right.+ \nonumber \\
& & \left.\left|\pm\right>\left<-s'\right| c_{-1\uparrow} + c^{\dagger}_{-1\uparrow} \left|-s'\right>\left<\pm\right|   \right) 
\end{eqnarray}
where $t$ is the lead-molecule tunneling amplitude. Note that the electron tunneling involves either spin up electrons or spin down electrons hopping from the leads into the single molecule magnet (see Fig.(\ref{scheme})).
If we substitute Eq.(\ref{eqsym}) into the Hamiltonian in Eq. (\ref{eq:SMM-lead}) we obtain the following effective Hamiltonian for the interaction of the SMM with the leads 
\begin{eqnarray}
\label{eq:SMM-lead_effective}
{\cal H}_{\rm SMM-lead}&=& -\frac{t}{N} \left( 
\left|-s\right>\left<-s'\right| c_{1\downarrow} + c^{\dagger}_{1\downarrow} \left|-s'\right>\left<-s\right| \right. +\nonumber \\
& & \left.\left|-s\right>\left<-s'\right| c_{1\uparrow} + c^{\dagger}_{1\uparrow} \left|-s'\right>\left<-s\right|  \right.+\nonumber \\
& & \left. 
\left|-s\right>\left<-s'\right| c_{-1\downarrow} + c^{\dagger}_{-1\downarrow} \left|-s'\right>\left<-s\right| \right.+ \nonumber \\
& & \left.\left|-s\right>\left<-s'\right| c_{-1\uparrow} + c^{\dagger}_{-1\uparrow} \left|-s'\right>\left<-s\right|   \right) \nonumber \\
& & -\Lambda \left( 
\left|s\right>\left<-s'\right| c_{1\downarrow} + c^{\dagger}_{1\downarrow} \left|-s'\right>\left<s\right| +\right. \nonumber \\
& & \left.\left|s\right>\left<-s'\right| c_{1\uparrow} + c^{\dagger}_{1\uparrow} \left|-s'\right>\left<s\right| +\right.  \nonumber \\
& & \left.\left|s\right>\left<-s'\right| c_{-1\downarrow} + c^{\dagger}_{-1\downarrow} \left|-s'\right>\left<s\right| +\right. \nonumber \\
& & \left.\left|s\right>\left<-s'\right| c_{-1\uparrow} + c^{\dagger}_{-1\uparrow} \left|-s'\right>\left<s\right|\right)
\end{eqnarray}
where $\Lambda=t\lambda_{s,-s}(N_--N_+)/(N_+N_-)$ and
\begin{equation}
\frac{1}{N}=\left[\frac{-\Delta_{s,-s}+\sqrt{\Delta^2_{s,-s}+\lambda^2_{s,-s}}}{N_+}+\frac{\Delta_{s,-s}+\sqrt{\Delta^2_{s,-s}+\lambda^2_{s,-s}}}{N_-}\right].
\label{sflip}
\end{equation}
The first term in equation (\ref{eq:SMM-lead_effective}) represents the electron tunneling from the lead to the SMM without spin-flip processes with tunneling amplitude $t/N$ and the second term respresents the electron tunneling from the lead to the SMM with spin-flip processes with tunneling amplitude $\Lambda$. Interestingly, if $\Delta_{s,-s}\rightarrow 0$ or $\lambda_{s,-s}\rightarrow 0$ then $\Lambda\rightarrow 0$ and there is no spin-flip processes. This shows that magnetic quantum coherence and the coupling between the states $|\pm s\rangle$ opens a new channel for the electron to tunnel through the molecular transistor which interferes with the direct channel. \\
\section{Results}
\label{sec:result}
\begin{figure}[!ht]
  \begin{center}
    \begin{tabular}{cc}
      \resizebox{42mm}{!}{\includegraphics{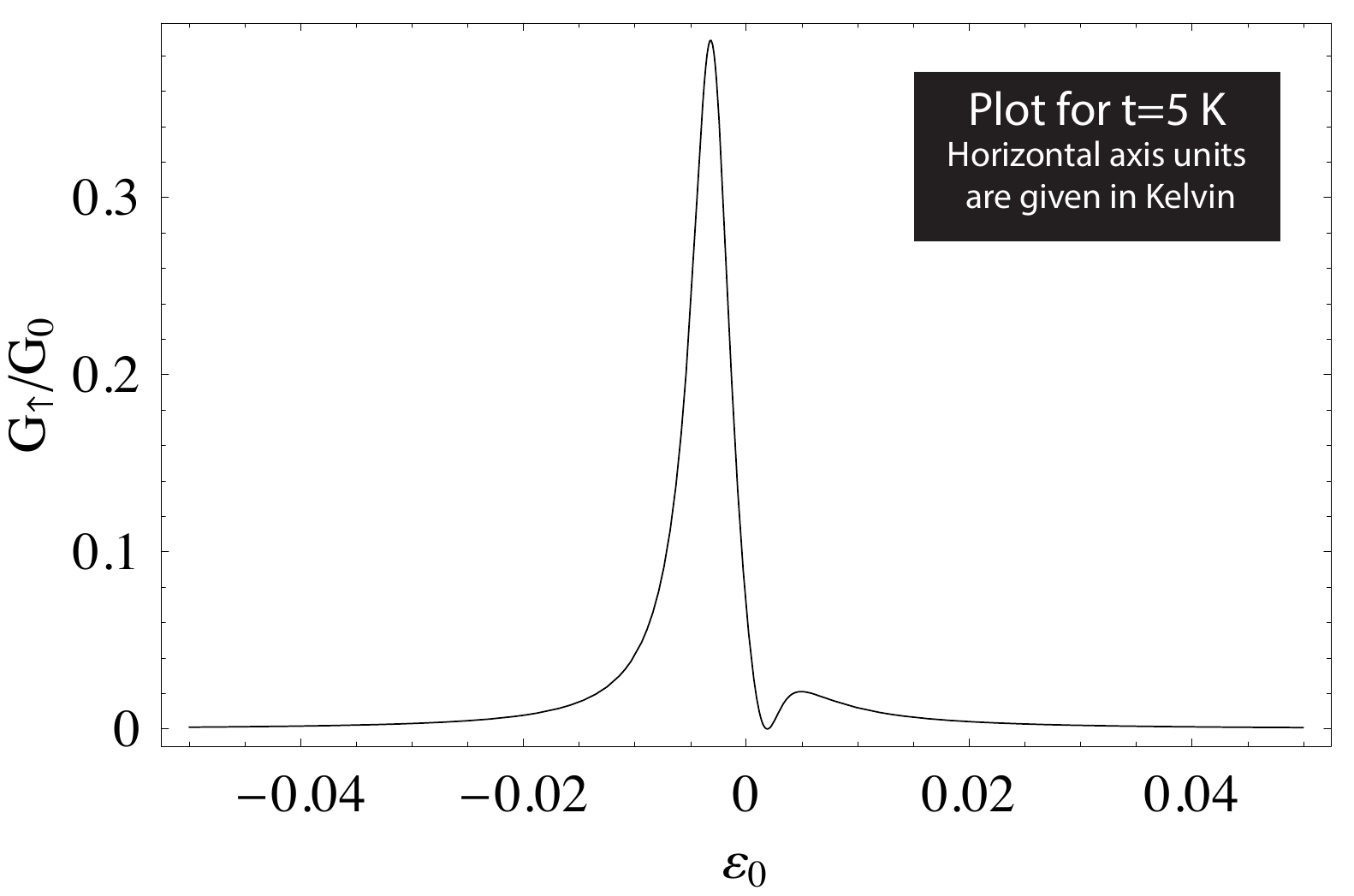}} &
      \resizebox{42mm}{!}{\includegraphics{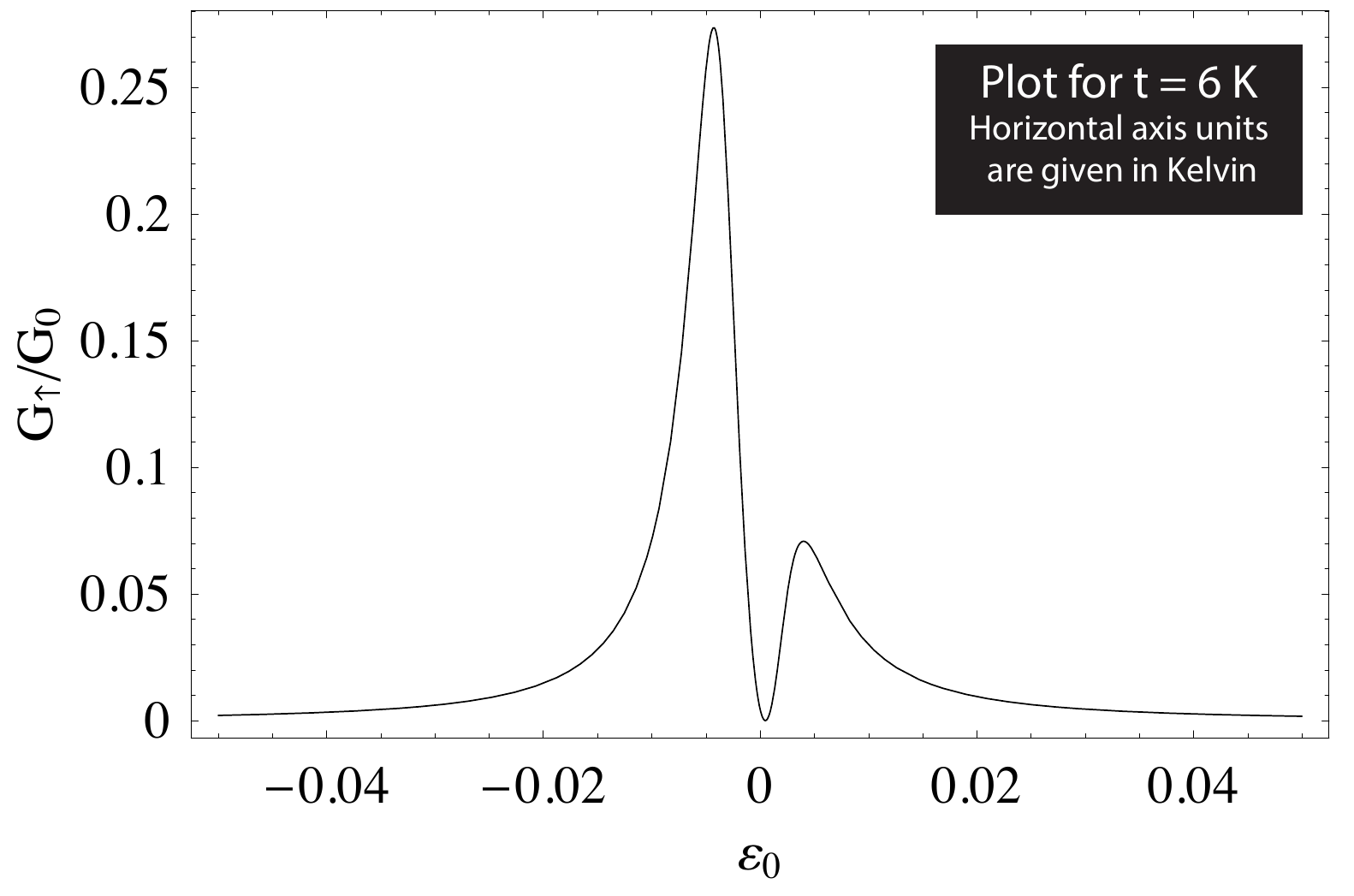}} \\ 
    \multicolumn{1}{c}{\mbox{\bf (a)}} &
		\multicolumn{1}{c}{\mbox{\bf (b)}} \\ 
      \resizebox{42mm}{!}{\includegraphics{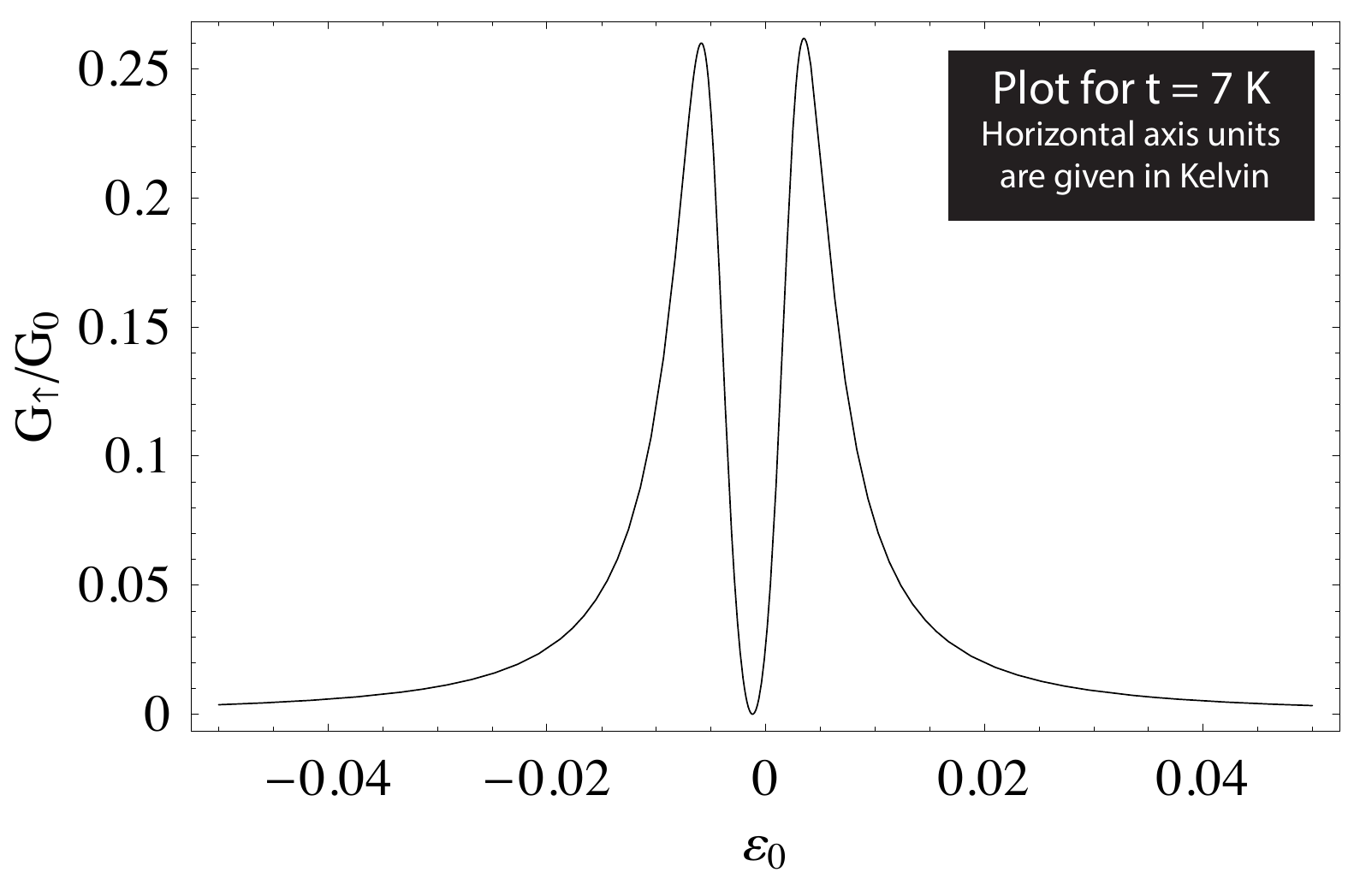}} &
      \resizebox{42mm}{!}{\includegraphics{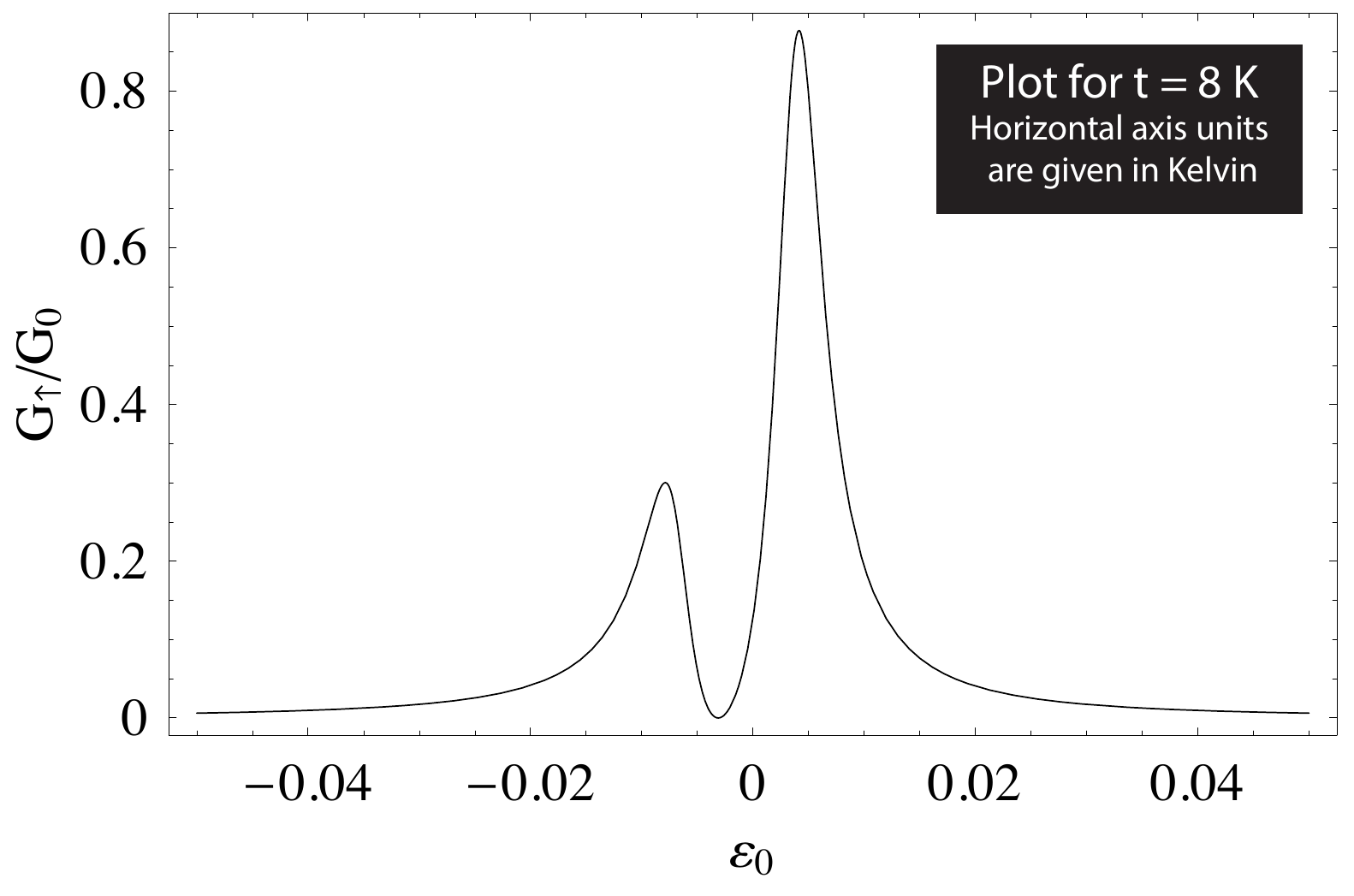}} \\
    \multicolumn{1}{c}{\mbox{\bf (c)}} &
		\multicolumn{1}{c}{\mbox{\bf (d)}} 
    \end{tabular}
    \caption{Plots showing the spin-dependent conductance as a function of the gate voltage for different values of the lead-molecule tunneling amplitude, (a) $t=5$ K(b)
    $t=6$ K, (c) $t=7$ K and (d) $t=8$ K, with a constant magnetic field of $h_{\parallel}=0.25$ mK. Note how the MQC effect increases with the coupling interaction.}
	\label{Conduc}
  \end{center}
\end{figure}

In what follows we present our results for the spin dependent conductance.
The stationary state of the total Hamiltonian can be written as \cite{ore}
\begin{equation}
|\psi_{\sigma,\pm s}\rangle=\sum_{j=-\infty,j\neq0}^{\infty}a_{j\sigma,\pm s'}|j\rangle|\pm s'\rangle+b_{\sigma,\pm s}|0\rangle|\pm s\rangle,
\label{eqstatio}
\end{equation}
where $a_{j\sigma,\pm s'}$ and $b_{\sigma,\pm s}$ are the probability amplitudes to find the electron at the site $j\ne 0$ or at the SMM site $j=0$, respectively, with energy $\omega=\epsilon_{\sigma,-s}-2v\cos(k)$ or $\omega=\epsilon_{\sigma,s}-2v\cosh(\kappa)$, where $\epsilon_{\sigma,\pm s}=\Delta_z[\sigma_z]_{\sigma\sigma}/2$. Substituting Eq. (\ref{eqstatio}) into the Hamiltonian we obtain the following linear difference equations:
\begin{eqnarray}
\label{diffeq}
\mbox{for $j\neq -1,0,1$ we have} \\ \nonumber
(\omega-\epsilon_{\sigma,\pm s})a_{j\sigma,\pm s'}&=&-v(a_{(j+1) \sigma,\pm s'}+a_{(j-1) \sigma,\pm s'}),\\ \nonumber
\mbox{for $j=-1,0,1$ we have} \\ \nonumber
(\omega-\epsilon_{\sigma,\pm s})a_{-1\sigma,\pm s'}&=&-va_{-2\sigma,\pm s'}-\frac{t}{N} b_{\sigma,\pm s}-\Lambda b_{\sigma,\mp s}, \\ \nonumber 
(\omega-\epsilon_{\sigma,\pm s})a_{1\sigma,\pm s'}&=&-va_{2\sigma,\pm s'}-\frac{t}{N} b_{\sigma,\pm s}-\Lambda b_{\sigma,\mp s}, \\ \nonumber  
(\omega-\tilde{\varepsilon}_{0\sigma,\pm s})b_{\pm s}&=&-\frac{t}{N}(a_{1\sigma,\pm s'}+a_{-1\sigma,\pm s'})+\\ \nonumber & &   
-\Lambda (a_{1\sigma,\mp s'}+a_{-1\sigma,\mp s'}).
\end{eqnarray}
where $\tilde{\varepsilon}_{0\sigma,\pm s}=\varepsilon_0+\lambda_{s,-s}$.\\
In order to obtain the solution of the above equations we assumed that the spin-up electrons are described by plane waves with unitary incident amplitude coming from the left, with $r$ and $\tau$ being the reflection and transmission amplitudes, therefore
\begin{eqnarray} 
\label{eqwave}
a_{j\uparrow,-s'}=e^{ikj}+re^{-ikj}, \quad j<0, \\ \nonumber
a_{j\uparrow,-s'}=\tau e^{ikj}, \quad j>0, \\ \nonumber
a_{j\downarrow,s'}=Ae^{\kappa j}, \quad j<0 \\ \nonumber
a_{j\downarrow,s'}=Be^{-\kappa j}, \quad j>0.
\end{eqnarray}
Substituting Eq. (\ref{eqwave}) into Eq. (\ref{diffeq}), we get and inhomogeneous system of linear equations for the unknowns $A$, $B$, $r$ and $\tau$, leading to the following expression for the transmission amplitude 
\begin{widetext}
\begin{eqnarray}
\tau=\frac{2i\alpha\sin(k)\left[t^2(\omega-\tilde{\varepsilon}_{0\downarrow})/N^2-(\omega-\tilde{\varepsilon}_{0\uparrow})\Lambda^2-2\alpha^2ve^{-\kappa}\right]}{[t^2(\omega-\tilde{\varepsilon}_{0\uparrow}+2\alpha e^{ik})(\omega-\tilde{\varepsilon}_{0\downarrow}-2\alpha e^{-\kappa})/N^2+\Lambda^2(\omega-\tilde{\varepsilon}_{0\downarrow}+2\alpha e^{ik})(-\omega+\tilde{\varepsilon}_{0\uparrow}+2\alpha e^{-\kappa})]}
\label{trans}
\end{eqnarray}
\end{widetext}
where $\alpha=t^2(2-N^2)/vN^2$. If $\lambda_{s,-s}=0$ the above equation reduces to a single resonance
\begin{equation}
\tau=\frac{i\Gamma}{(\omega-\varepsilon_{0})+i\Gamma}
\label{eqsin}
\end{equation}
where $\Gamma=2t^2\sin(k)$ is the width of the resonance centered at $\varepsilon_{0}$.
The conductance for the electron tunneling across the SMM transistor is calculated by means of the Landauer formalism at zero temperature, i.e.
\begin{equation}
G=\frac{e^2}{h}T(E_F),
\label{eqlan}
\end{equation}
where $e$ is the charge unit, $h$ is the Planck constant, $E_F$ is the chemical potential and $T=|\tau|^2$. The transmission probability at $\omega=E_F=0$ for the spin dependent conductance is
\begin{widetext}
\begin{equation}
G_{\uparrow}=\frac{e^2}{h}\frac{4\sin^2(k_F)[\alpha (\varepsilon_0+\xi_-)-t^2\sqrt{(\mu_BgH_z)^2+\Delta_{s,-s}^2}/2]^2}{|(\varepsilon_0-\xi_-)(\varepsilon_0-\xi_+)+2t^2\sqrt{(\mu_BgH_z)^2+\Delta_{s,-s}^2}(e^{ik_F}+e^{-\kappa_F})/2-((\mu_BgH_z)^2+\Delta_{s,-s}^2)/4|^2}
\label{spd}
\end{equation}
\end{widetext}
where $\xi_+=2\alpha\cos(k_F)$, $\xi_-=2t^2e^{-\kappa_F}$, $k_F=\cos^{-1}(1-\frac{\Delta_z}{4v})$ and $\kappa_F=\cosh^{-1}(1+\Delta_z/4v)$. In all subsequent calculations we will express energy in temperature units. Therefore, a magnetic field of $H_z=1$ T is equal to $h_{\parallel}=1.34$ K and $1$ K is equal to a frequency of $\omega=1.31\times10^{11}$ Hz.
In what follows we present our results for $\Delta_z/4v=0.1$ and at a fixed longitudinal magnetic field of $h_{\parallel}=0.25$ mK. We used $\Delta_{s,-s}=0.005$ K, $\lambda_{s,-s}=0.05$ K and $N=1.01$  for our calculations. The temperature limit to observe this effect would be $T<<0.01$ K.
In Figure (\ref{Conduc}) the spin-dependent conductance versus the gate voltage is depicted for different values of the lead-molecule tunneling amplitude. The conductance shows two resonances as a function of the Fermi energy.  
The strong dependence of the conductivity with the tunneling amplitude follows directly from the effective Hamiltonian given in equation (\ref{eq:SMM-lead_effective}) in which the electron tunneling with and without spin-flip processes are proportional to the tunneling amplitude. If the lead-molecule tunneling amplitude is weak then a small fraction of the electrons will have a spin-flip process and the interference between different tunneling trajectories will not be strong. On the other hand, as the lead-molecule tunneling amplitude increases the interference between different paths is enhanced, therefore, the shift in the peaks is due to the spin-flip processes in the electron transmission. This result shows that the conductance through the SMM transistor depends on the MQC. 
    
\section{Conclusions}
We have investigated electron transport through a Ni$_4$ molecular transistor with Zeeman spin splitting and magnetic quantum coherence. We have shown that the transport through the SMM transistor presents a quantum interference between different tunneling trajectories due to the magnetic quantum coherence. Our results are in contrast to the system analyzed in Ref. \onlinecite{gabriel} where it is essential to have oppositely spin polarized leads and incoherent spin states in order for quantum interference to take place. This results provide a new method to observe the coherent spin tunneling through MQC resonances of the spin dependent conductance as a function of the gate voltage and of the magnetic field applied in the single molecule magnet.

\end{document}